\documentstyle[prl,aps,epsfig,multicol]{revtex}
\begin{document}
\draft
\newcommand{\bet}{$\beta^{ \prime \prime} - (ET)_2SF_5CH_2CF_2SO_3$}
\newcommand{\bret}{$\kappa-(ET)_2 Cu \left[N \left( CN \right)_2\right]Br$}
\newcommand{\set}{$\kappa-(ET)_2Cu \left( NCS \right)_2$}
\newcommand{\dlt}{$\Delta \lambda (T)$}
\newcommand{\dltpar}{$\Delta \lambda_{\parallel} (T)$}
\newcommand{\ltpar}{$\lambda_{\parallel} (T)$}
\newcommand{\dltperp}{$\Delta \lambda_{\perp} (T)$}
\newcommand{\ltperp}{$\lambda_{\perp} (T)$}

\title{Unusual temperature dependence of the London penetration depth in all-organic
\bet~single crystals}

\author{R.~Prozorov and R.~W.~Giannetta}
\address{Loomis Laboratory of Physics, University of Illinois at
Urbana-Champaign, 1110 West Green St., Urbana, Illinois 61801.}

\author{J.~Schlueter and A.~M.~Kini}
\address{Chemistry and Materials Science Division, Argonne National
Laboratory, Argonne, Illinois 60439.}

\author{J.~Mohtasham, R.~W.~Winter, G.~L.~Gard}
\address{Department of Chemistry, Portland State University, Portland,
Oregon 97207.}

\date{\today}
\maketitle

\begin{abstract}
The temperature dependence of the in-plane, \ltpar, and interplane, \ltperp,
London penetration depth was measured in the metal-free all-organic
superconductor \bet~($T_c \approx$ 5.2 K). \dltpar~$\propto T^3$ up to 0.5
$T_c$, a power law previously observed only in materials thought to be $p-$wave
superconductors. $\lambda_{\perp}$ is larger than the sample dimensions down to
the lowest temperatures (0.35 K), implying an anisotropy of
$\lambda_{\perp}/\lambda_{\parallel} \approx 400-800$.
\end{abstract}

\pacs{74.70.Kn, 74.25.Nf}

\begin{multicols}{2}
\narrowtext

Despite intensive study, neither the pairing mechanism nor the symmetry of the
order parameter has been conclusively established in organic superconductors of
the $\kappa-(BEDT-TTF)_2X$ class. (Henceforth ``BEDT-TTF`` will be abbreviated
by "ET".)  For the most thoroughly investigated materials, \set~($T_c \approx$
9.5 K) and \bret~($T_c \approx$ 12 K), there is some evidence for a $d-$wave
pairing \cite{dwave,carrington99}. However, recent penetration depth
measurements revealed an unusual fractional power law variation, \dlt $\propto
T^{3/2}$, unlike that of any other superconductor \cite{carrington99}. While
this exponent is consistent with a novel three-fluid model \cite{kosztin98}, it
is also suggestive of a magnetic excitation. In this paper we report
penetration depth measurements in \bet, a recently synthesized all-organic
superconductor free of metallic ions and in which magnetism is likely to be
negligible. This material is a strongly two dimensional, extreme type II
superconductor with $T_c \approx 5.2 K$.  It is metallic between 10 and 150 K
and semiconducting from 150 and up to 410 K \cite{wang99}. The upper critical
field parallel to the conducting planes exceeds the Pauli limit by 18\% raising
the possibility of either an inhomogeneous pairing state \cite{zuo99,fulde64}
or spin triplet order parameter \cite{lee97}. We determine the London
penetration depth for supercurrents both along ($\lambda_{\parallel}$) and
perpendicular ($\lambda_{\perp}$) to the conducting planes. The penetration
depth is extremely anisotropic, with $\lambda_{\perp}$ roughly 800 times larger
than $\lambda_{\parallel}$. Notably, $\lambda_{\parallel} \propto T^3$ which
might imply an energy gap with nodes, but is difficult to reconcile with either
$p$ or $d-$wave models in two dimensions. We suggest that this power law may
arise from the unusual phonon spectrum in this material.

Single crystals of \bet~were grown at Argonne National Laboratory by an
electrocrystallization technique described elsewhere \cite{geiser96}. The high
conductance layers correspond to the $ab$ plane and the $c^*$ axis is normal to
the planes. This designation is similar to cuprates, while different from the
$\kappa-(ET)_2X$ materials. The room-temperature interplane resistivity is
roughly 700 $\Omega~cm$ while the in-plane resistivity is about 0.2 $\Omega~cm$
\cite{su99}. Two crystals - $0.5 \times 0.5 \times 0.3~mm^3$ and $0.8 \times
0.6 \times 0.3~mm^3$ were used for measurements. Each had a transition
temperature of approximately 5.2 K. A third crystal was used to measure the
absolute penetration depth. The penetration depth was measured with an 11 MHz
tunnel-diode driven LC resonator \cite{prozorov00}. Samples were mounted on a
movable sapphire stage with temperature controllable from 0.35 K to 50 K. The
low noise level, $\Delta f_{min}/f_{0}\approx 5 \times 10^{-10}$, resulted in a
sensitivity of $\Delta \lambda \leq 0.5$ \AA\ for our samples. An rf field was
applied either perpendicular to the conducting planes to probe \dltpar~or along
the $a-$axis to probe \dltperp.

The resonator frequency shift due to superconducting sample, $\Delta f \equiv
f(T)-f_0$, is given by \cite{prozorov00}:
\begin{equation}
\frac{\Delta f}{f_0}=\frac{V_{s}}{2V_{0} \left( 1- N \right)}\left( 1-
\frac{\lambda }{R} \tanh{\frac{R}{\lambda}} \right) \label{df}
\end{equation}
\noindent where $f_0$ is the frequency in the absence of a sample, $V_{s}$ is
the sample volume, $V_{0}$ is the effective coil volume and $N$ is the
effective demagnetization factor. The apparatus and sample - dependent constant
$\Delta f_{0} \equiv V_{s}f_0/(2V_{0} \left( 1- N \right) )$ was measured by
removing the sample from the coil in situ \cite{prozorov00}. For $\lambda \ll
R$, $\tanh{R/ \lambda} \approx 1$ and the change in $\lambda $ with respect to
its value at low temperature is $\Delta \lambda = -\delta f R/\Delta f_{0}$,
where $\delta f \equiv \Delta f(T) - \Delta f(T_{min})$. In the parallel
orientation ($H \parallel ab$), however, we had to use the full expression,
Eq.~(\ref{df}) to estimate $\lambda_{\perp}$ due to the weak screening in that
direction.

\begin{figure} [t]
\centerline{\psfig{figure=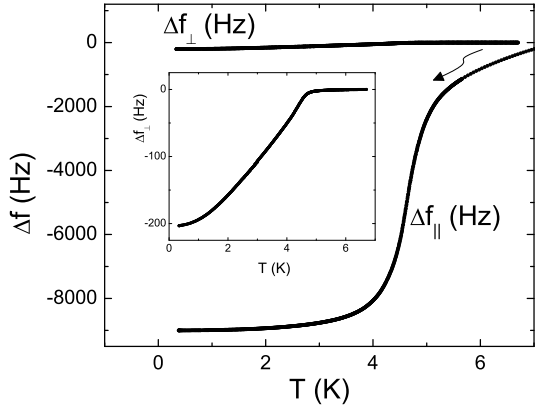,width=8.5cm}} \caption{Frequency variation
in parallel ($\Delta f_{\perp}$) and perpendicular ($\Delta f_{\parallel}$)
orientations of the magnetic with respect to superconducting layers. Usual
notation in terms of current flow is used. \textit{Inset:} zoom of $\Delta
f_{\perp} (T)$. Note substantial difference in shielding ability for two
orientations.} \label{cvsab}
\end{figure}

Figure \ref{cvsab} shows the frequency variation measured in two orientations
for zero DC magnetic field. For ($H \parallel ab$) the rf screening is
controlled by \ltperp~and is much weaker than in the ($H \parallel c^*$)
orientation, where the relevant screening length is \ltpar. Since all three
crystal dimensions were roughly comparable, this indicates that
$\lambda_{\perp}$ is several hundred times larger than $\lambda_{\parallel}$.
The inset shows an expanded view of the $\Delta f_{\perp}(T)$ curve. From the
total frequency variation and using Eq.~(\ref{df}) we estimate
$\lambda_{\perp}(0) \approx 800$ $\mu m$.

\begin{figure} [t]
 \centerline{\psfig{figure=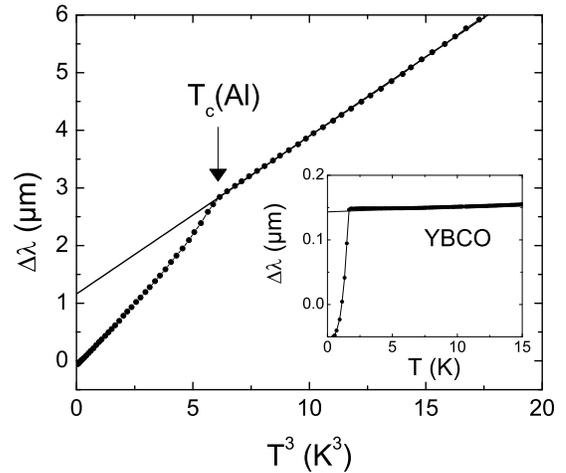,width=8.5cm}}
\caption{Measurements of the absolute value of $\lambda_{\parallel}(0)$ in
\bet. \textit{Inset:} Same technique applied to YBCO.} \label{al}
\end{figure}

To date, there have been no reported measurements of the zero temperature
penetration depth, $\lambda_{\parallel}(T = 0)$. We recently developed a new
method to determine $\lambda_{\parallel}(T = 0)$ that relies upon the change in
screening of an Al-coated sample as the temperature is reduced from above
$T_c(Al)$ to below $T_c(Al)$ \cite{prozorov00c}. The inset to Fig.~\ref{al}
shows the data obtained in a single crystal of YBCO. The method yields a value
of $0.145 \pm 0.01 \mu m$ which is within 5\% of literature values. The
mainframe of Fig.~\ref{al} shows the method applied to \bet. Since $T_c$ of
this material is only 5.2 K, its penetration depth is still changing at 0.35 K
and the method is less reliable than for cuprate superconductors. We estimate a
value of $\lambda_{\parallel}(T=0) = 1-2~\mu m$, in rough agreement with values
for other $ET$ compounds \cite{carrington99}, and leading to an anisotropy of
400 - 800. Our measurements provide only the average of \dltpar. Microwave
conductivity measurements revealed a small in-plane anisotropy of approximately
1.35 with a maximum along the $b$ axis \cite{wang99}.

\begin{figure} [t]
 \centerline{\psfig{figure=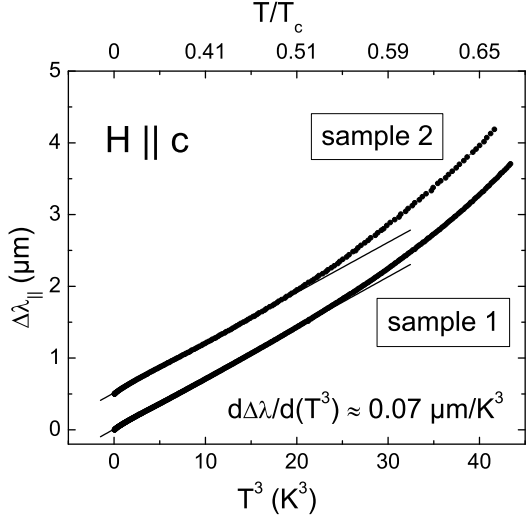,width=8.5cm}}
\caption{\dltpar~measured in two different crystals. (Data for sample 2 is
offset for clarity). Solid lines show fits to $T^3$ power law.} \label{twosamp}
\end{figure}

\begin{figure} [t]
 \centerline{\psfig{figure=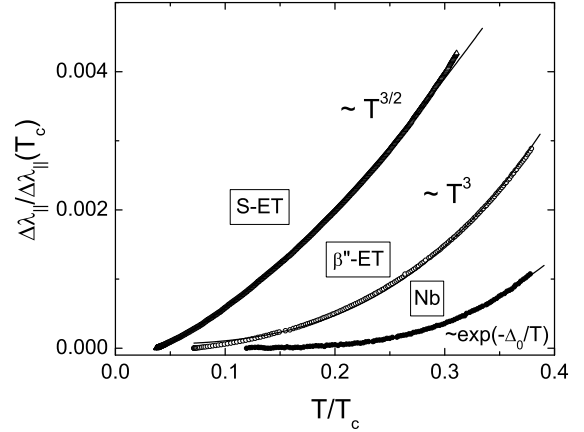,width=8.5cm}}
\caption{Comparison of the temperature variation of \dltpar~in different
systems. From bottom to top: polycrystalline Nb (solid line is a fit to a
standard weak coupling s-wave BCS low-temperature expansion with $\Delta
(0)/T_c=1.76$); \bet~(solid line is a fit to $T^3$); \set~(solid line is a fit
to $T^{3/2}$).} \label{compare}
\end{figure}

Figure~\ref{twosamp} shows the low temperature variation of \dltpar\ observed
in two samples of \bet. Data for sample 2 is offset for clarity. The horizontal
axis is $T^3$ showing that \dlt~$\propto T^3$ with a slope of 0.07 $\mu m/K^3$.
The cubic power law is obeyed up to $\sim T_c/2$. The $Al$ coated sample, shown
in Fig.~\ref{al} also showed \dlt~$\propto T^3$, but below $T_c(Al)$ the signal
from \bet~is screened by the Al coating. Both the n = 3 exponent and the wide
range over which it holds are unusual and have not been observed in cuprate
superconductors. To highlight the differences among superconductors, we plot in
Fig.~\ref{compare} the normalized low temperature variation of the penetration
depth in \set~(uppermost curve), \bet~(middle curve) and polycrystalline Nb for
comparison. Solid lines are the fits to $T^{3/2}$, $T^{3}$ and $\sqrt{\pi
\Delta (0)/2T}\exp{(-\Delta (0)/T)}$ variations. All data were taken in the
same apparatus.

It is possible that \bet~has an extremely anisotropic s-wave order parameter
and the $T^3$ variation is an effective, intermediate temperature power law
that only holds above the low temperature, exponential region. Our numerical
calculations show that anisotropic s-wave states, at least in weak coupling, do
not exhibit a $T^3$ variation over any extended range. In fact, the data in
Fig.~\ref{twosamp} shows a slight \textit{downward} deviation from $T^3$ at the
lowest temperatures, implying a decrease in the exponent - just the opposite of
exponential suppression. Strictly speaking, it is the power law variation of
the superfluid density $\rho_s$ which is most directly related to the structure
of the gap. \dltpar~is the measured quantity and its temperature variation only
asymptotically approaches that of $\rho_s$. The superfluid density versus
temperature was calculated from \dltpar~for $\lambda_{\parallel}(0) =
0.5,1,2,5~\mu m$. In each case, we found that a cubic power law remained the
best fit, although the range over which it held was reduced for smaller choices
of $\lambda_{\parallel}(0)$.

It is also possible that a small tilt of the $c*$ axis relative to the field
may induce interplane supercurrents and create an admixture of both \ltpar~and
\ltperp~in the data. If the applied field is tilted by $\theta$ relative to the
$c*-$axis the additional contribution to the observed frequency shift is given
by \cite{prozorov00},

\begin{equation}
\Delta f_{tilt} = \frac{f_0 V_{s}}{2V_{0}}\left( {1 - \left[ {\frac{{\lambda
_{\parallel} }}{d} + \frac{{\lambda_{\perp}}}{w}} \right]} \right)\sin ^2
\left( \theta \right) \label{tilt}
\end{equation}

The alignment was checked at room temperature by repeatedly attaching a sample
to the sapphire rod with vacuum grease and measuring the divergence of a laser
beam reflected off the sample surface. The average alignment error was never
more than 2 degrees. To be conservative, we consider a misalignment of 5
degrees and using the data for \dltperp~from Fig.~\ref{cvsab}, calculate a
maximum misalignment error of 4~\% in our determination of \dltpar~versus
temperature. This value is too small to change our conclusion about the
presence of an n = 3 exponent.

A $T^3$ variation of $\lambda_{\parallel}$ is unusual, but was predicted for a
three dimensional $p-$wave superconductor with an equatorial line of nodes: the
so-called polar state with $\Delta (\hat{k})=\Delta_0(T) \hat{k} \cdot \hat{l}$
\cite{gross86,einzel86}. Here, $\hat{l}$ is the axis of gap symmetry which must
lie parallel to the vector potential $\overrightarrow{A}$ in order to obtain a
cubic power law. If $\hat{l}$ is perpendicular to $\overrightarrow{A}$ the
dependence is linear in T. The relevance to our data is questionable since
\bet~is strongly two dimensional and both $d$ and $p-$ wave states must have
line nodes perpendicular to the $ab$ plane, giving a linear $T$ dependence. A
$T^3$ dependence would then require an angular variation of the gap near the
node, $\Delta(\phi) \propto \phi^{1/3}$, for which there is no obvious
justification. Previous tunnel diode measurements of the penetration depth in
$UPt_3$, believed by many to be a p-wave superconductor, revealed intermediate
exponents ranging from n = 2-4 depending upon surface preparation
\cite{signore95}.
\noindent However, lower frequency measurements on the same samples gave lower
power laws (n = 1-2) for reasons not understood, but possibly related to
surface dissipation. SQUID measurements of the penetration depth in the heavy
fermion material $UBe_{13}$ gave n = 2, which could arise either from point
nodes or impurity scattering \cite{gross86,einzel86,hirschfeld93}. The latter
might be an issue in \bet~since, at the low end, our data show a slight
tendency toward a lower power law, possibly $n = 2$. Recent measurements in
$Sr_2RuO_4$, also thought to be $p-$wave superconductor, have shown $\lambda
\approx\ T^3$ in one sample, attributed to a combination of impurity scattering
and nonlocality in a superconductor with line nodes \cite{bonalde00}. \bet~is
an extreme type II material and nonlocality is unlikely to be an issue until
one reaches temperatures of order $(\xi/\lambda) T_c \approx 0.05$ K
\cite{kosztin96}. Finally, on general grounds p-wave pairing is favored in
materials with a tendency toward ferromagnetism, for which there is no evidence
in this material. Although the discovery of a new pairing symmetry is
appealing, \bet~is sufficiently complex that other possibilities should be
considered. Recent heat capacity measurements suggest a strong-coupling s-wave
BCS state. They also indicate the presence of optical modes in the 20-40 K
energy range \cite{wanka98}. Some time ago, it was shown theoretically that the
coupling of electrons to low frequency, localized vibrations can give a
temperature dependence to the effective mass and thus a power law to the London
penetration depth over and above that due to the superfluid fraction
\cite{klimovitch93}. For example, a phonon density of states $g(E)$ varying as
$E^2$ may give rise to a $T^3$ power law for an s-wave superconductor, in the
absence of vertex corrections. Under most circumstances vertex corrections
raise the power to $T^5$ making the effect extremely small, but this may not be
true here.  Our data suggest that strong coupling calculations involving a
realistic phonon spectrum may be relevant for organic superconductors.  We also
wish to stress the desirability of NMR measurements in \bet~to help determine
the parity of the order parameter.

\begin{acknowledgments} We wish to thank M.~B.~Salamon for useful discussions and for
providing results on $Sr_2RuO_4$ prior to publication. Research at Urbana was
supported through State of Illinois ICR funds. Research at Argonne was
supported by the U.S. Department of Energy, Office of Basic Energy Sciences,
Division of Materials Sciences, under contract No. W-31-109-ENG-38. Research at
Portland State University was supported by NSF grant No. CHE-9904316 and the
Petroleum Research Fund, ACS-PRF 34624-AC7.
\end{acknowledgments}

\end{multicols}

\begin{references}

\bibitem{dwave} A. Kawamoto, K. Miyagawa, Y. Nakazawa, and K. Kanoda, Phys.
Rev. Lett. \textbf{74}, 3455 (1995); H. Mayaffre, P. Wzietek, D. Jerome,, C.
Lenoir, and P. Batail, Phys. Rev. lett. \textbf{75}, 4122 (1995); S. M. De
Soto, C. P. Slichter, A. M. Kini, H. H. Wang, U. Geiser, and J. M. Williams,
Phys. Rev. B, \textbf{52}, 10364 (1995); S. Belin, K. Behnina, and A. Deluzet,
Phys. Rev. Lett., \textbf{81} 4728 (1998).

\bibitem{carrington99} A. Carrington, I. J. Bonalde, R. Prozorov, R. W.
Giannetta, A. M. Kini, J. Schlueter, H. H. Wang, U. Geiser, and J. M. Williams,
Phys. Rev. Lett. \textbf{83}, 4172 (1999).

\bibitem{kosztin98} I. Kosztin, Q. J. Chen, B. Janko and K. Levin, Phys. Rev. B
\textbf{58}, R5936 (1998).

\bibitem{wang99} H. H. Wang, M. L. VanZile, J. A. Schlueter, U. Geiser, A.
M. Kini, P. P. Sche, H. J. Koo, M. H. Whangbo, P. G. Nixon, R. W. Winter, and
G. L. Gard, J. Phys. Chem. B \textbf{103}, 5493 (1999).

\bibitem{zuo99} F. Zuo, P. Zhang, X. Su, J. S. Brooks, J. A. Schlueter, J.
Mohtasham, R. W. Winter, and G. L. Gard, J. Low Temp. Phys. \textbf{117}, 1711
(1999).

\bibitem{fulde64} P. Fulde and R. A. Ferrel, Phys. Rev. \textbf{135}, A550 (1964);
A.I. Larkin and Yu.N. Ovchinnikov, Sov. Phys. JETP \textbf{20}, 762 (1965).

\bibitem{lee97} I. J. Lee, M.J. Naughton, G. M. Danner, P. M. Chaikin, Phys. Rev.
Lett. \textbf{78}, 3555 (1997).

\bibitem{geiser96} U. Geiser, J. A. Schluleter, H. H. Wang, A. M. Kini, J.
M. Williams, P. P. Sche, H. I. Zakowicz, M. L. VanZile, J. D. Dudek, J. Am.
Chem. Soc. \textbf{118}, 9996 (1996).

\bibitem{su99} X. Su, F. Zuo, J. A. Schlueter, J. M. Williams, P. G. Nixon,
R. W. Winter, and G. L. Gard, Phys. Rev. B: \textbf{59}, 4376 (1999).

\bibitem{prozorov00} R. Prozorov, R. W. Giannetta, A. Carrington, and F. M.
Araujo-Moreira, Physical Review B \textbf{62}, 115 (2000).

\bibitem{prozorov00c} R. Prozorov, R. W. Giannetta, A. Carrington, P.
Fournier, R. L. Greene, P. Guptasarma, D. G. Hinks, and A. R. Banks, Appl.
Phys. Lett., submitted; cond-mat/0007013.

\bibitem{gross86} F. Gross, B. S. Chandrasekhar, D. Einzel, K. Andres, P.
J. Hirschfeld, H. R. Ott, J. Beuers, Z. Fisk, and J. L. Smith, Z. Phys. B
\textbf{64}, 174 (1986).

\bibitem{einzel86} D. Einzel, P. J. Hirschfeld, F. Gross, B. S.
Chandrasekhar, K. Andres, H. R. Ott, J. Beuers, Z. Fisk, and J. L. Smith, Phys.
Rev. Lett. \textbf{56}, 2513 (1986).

\bibitem{signore95} P. J. C. Signore, B. Andraka, M. W. Meisel, S. E. Brown,
Z. Fisk, A. L. Giorgi, J. L. Smith, F. Gross-Alltag, E. A. Schuberth, A. A.
Menovsky, Phys. Rev. B. \textbf{52}, 4446 (1995).

\bibitem{hirschfeld93} P. J. Hirschfeld and N. Goldenfeld, Phys. Rev. B
\textbf{48}, 4219 (1993).

\bibitem{bonalde00}I. Bonalde, B. D. Yanoff, M. B. Salamon, D. J. Van Harlingen,
E. M. E. Chia, Z. Q. Mao, Y. Maeno (preprint).

\bibitem{kosztin96}I. Kosztin and A. J. Leggett, Phys. Rev. Lett. \textbf{79},
135(1997).

\bibitem{wanka98} S. Wanka, J. Hagel, D. Beckmann, J. Wosnitza, J. A. Schlueter,
J. M. Williams, P. G. Nixon, R. W. Winter, G. L. Gard, Phys. Rev. B
\textbf{57}, 3084 (1998).

\bibitem{klimovitch93} G. V. Klimovitch, JETP Lett. \textbf{59}, 786 (1993).

\end{references}
\end{document}